\documentclass[10pt,a4paper]{article}
\usepackage{graphicx,times,color,bm,amsmath,textcomp,url,ulem} 
\usepackage[affil-it]{authblk}

\newcommand{\equref}[1]{equation~(\ref{#1})}

\newcommand{\figref}[1]{Fig.~\ref{#1}}

\newcommand{\Fieq}{\ensuremath{F_{i}^{\mathrm{eq}}}}

\newcommand{\Fiav}{\ensuremath{F_{i}^{\mathrm{av}}}}

\newcommand{\bmv}{{\ensuremath{{\bm{v}}}}}

\newcommand{\bmr}{{\ensuremath{{\bm{r}}}}}

\newcommand{\bmc}{{\ensuremath{{\bm{c}}}}}

\newcommand{\bmi}{{\ensuremath{{\bm{i}}}}}

\newcommand{\bmj}{{\ensuremath{{\bm{j}}}}}

\newcommand{\bmS}{{\ensuremath{{\bm{S}}}}}

\definecolor{red4}{rgb}{0.6,0,0}

\definecolor{green4}{rgb}{0,0.6,0}

\definecolor{green4}{rgb}{0,0.6,0}

\definecolor{blue4}{rgb}{0,0,0.6}

\definecolor{bluegreen4}{rgb}{0,0.6,0.6}

\newcommand{\blue}[1]{{#1}}

\newcommand{\red}[1]{\textcolor{red}{#1}}

\newcommand{\referee}[1]{\textcolor{black}{#1}}

\newcommand{\varnik}[1]{\textcolor{black}{#1}}

\newcommand{\green}[1]{{#1}}

\newcommand{\hide}[1]{{}}

\newcommand{\newstrike}[1]{\sout{\red{}}}

\newcommand{\strike}[1]{{}}

\newcommand{\rond}[3]{\partial^{#1}_{#2}F^{#3}_{i}}

\newcommand{\rondDo}[4]{\partial^{#1}_{#2}\partial^{#1}_{#3} F^{#4}_{i}}

\newcommand{\F}[1]{F^{#1}_{i}}

\newcommand{\Hoda}[1]{\textcolor{black}{#1}}

\newcommand{\HodaRemove}[1]{}

\begin{document}

\title{Chapman-Enskog Analysis of Finite Volume Lattice Boltzmann Schemes}

\author[1,2]{Nima H. Siboni 
\thanks{Corresponding author: siboni@mpie.de}}
\author[2]{Dierk Raabe
\thanks{d.raabe@mpie.de}}
\author[2,3]{Fathollah Varnik
\thanks{fathollah.varnik@rub.de}}
\affil[1]{Aachen Institute for Computational Engineering and Sciences (AICES), RWTH-Aachen, Germany.}
\affil[2]{Max-Planck-Institute f\"{u}r Eisenforschung GmbH, D\"usseldorf, Germany.}
\affil[3]{Interdisciplinary Centre for Advanced Materials Simulation (ICAMS), Ruhr-Universit\"{a}t Bochum, Germany.}


\date{February 25, 2009}

\maketitle
\begin{abstract}

In this paper, \green{we provide } a systematic analysis of some finite volume lattice Boltzmann schemes\green{ in two dimensions. A complete iteration cycle in time evolution of} discretized distribution functions {is formally divided} into collision and propagation \green{(streaming)} steps. Considering mass and momentum conserving properties of the collision step, \green{it becomes obvious that} changes in the momentum of finite volume cells is just due to the propagation step. Details of the propagation step are discussed \green{for different approximate schemes for the evaluation of fluxes at the boundaries of the finite volume cells}. \green{Moreover,} a full Chapman-Enskog analysis is conducted \green{allowing} to recover \green{the} Navier-Stokes equation. \green{As an important result of this analysis, the relation between the lattice Boltzmann relaxation time and the kinematic viscosity of the fluid is derived for each approximate flux evaluation scheme. In particular, it is found that the constant upwind scheme leads to a positive numerical viscosity while the central scheme as well as the linear upwind scheme are free of this artifact.}

\end{abstract}
\section{Introduction}
In recent years, mesoscopic methods such as stochastic rotation dynamics (SRD)~\cite{malevanets,Ihle}, dissipative particle dynamics (DPD)~\cite{DPD,DPD2} and the lattice Boltzmann method \blue{(LBM)}~\cite{Mcnamara,Chen,Higuera89,Qian,Benzi92} appeared as alternatives to conventional computational fluid dynamic (CFD) methods for the simulation of several complex fluid dynamical problems \varnik{ranging from
two-phase flow through porous media \cite{Gunstensen}, particle-fluid suspensions \cite{Ladd2001} and high Reynolds number flows \cite{Succi2002,Varnik2006,Varnik2007a,Varnik2007b}.}

\begin{figure}[h!]
\centering
\includegraphics[ width=0.3\textwidth ]{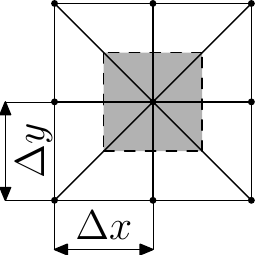}
\caption{Uniform mesh \blue{with} $\Delta x=\Delta y$ and the \blue{finite volume} cell constructed on it (filled with gray). \blue{The filled circles are lattice Boltzmann nodes. The time evolution of the population densities at the central node is controlled by the projected (onto the surface normal) sum of all the fluxes entering/leaving the gray region.}}
\label{CV}
\end{figure}

The original lattice Boltzmann method, proposed as\strike{floating-number} \blue{probability density based (and hence a coarse grained)} counterpart of lattice-gas cellular automata \cite{Frisch,Wolfram,Frisch2}, attracted much interest to be a practical computational fluid dynamics tool. However, one of the major drawbacks of the original lattice Boltzmann method was its\strike{constrain} \blue{restriction} to uniform\strike{space-time} lattices, leading to problems\strike{while} \blue{when} dealing with practical CFD applications. In order to overcome this limitation, many efforts were\strike{done} \blue{undertaken} to enhance LBM's efficiency when dealing with complex geometries or multi scale problems which require non-uniform or unstructured grids. These\strike{methods} \blue{extensions of the standard LBM} are generally\strike{adopted from} \blue{motivated by the} conventional CFD methods. Among the most important techniques are local grid refinement \cite{Filippova,Chopard,Yu}, interpolation supplemented finite difference method \cite{He98,Gu2003} \blue{as well as} various types of finite volume formulations \cite{Succi,Nanneli,UbertiniSucci,Peng,Xi_PRE}. These different approaches extent the lattice Boltzmann method to non-uniform and even unstructured grids.
Among these methods, finite volume approaches are of\strike{most} \blue{a rather} simple and flexible\strike{choice} \blue{form. As the name suggests, in a finite volume approach, the space is divided into a number of subvolumes, each containing a finite (but not necessarily equal) fraction of the total available volume of the computational domain. When applying to a space spanned by lattice nodes (as is the case in the lattice Boltzmann method), \HodaRemove{this}\Hoda{the} division of the space is done in such a way as to ensure that each subvolume contains (surrounds) one and only one lattice node. The time evolution (update) of the population densities assigned to a given lattice node is then governed by the net mass current into the subvolume containing that node. Obviously, the net mass flux into a subvolume is computed as the surface integral of the mass flux vector projected onto the direction normal to the surface surrounding the node under consideration.}
Although finite volume formulations are used extensively, there is still lack of analytical analysis of this type of LBM modifications. There are\blue{, however,} some theoretical arguments, based on the inspection of the dispersion relation associated with the unstructured lattice Boltzmann scheme \cite{UbertiniSucci}. More recently, through establishing a link between the lattice Boltzmann scheme and the finite volume method, general relations that define mass and momentum fluxes is proposed and the treatment of the boundary condition is studied \cite{DuboisLallemand}. Here we will perform a full Chapman-Enskog analysis of some typical finite volume approaches to introduce a framework for further analytical investigations.\strike{In this paper, a uniform mesh is chosen to show} \blue{In order to keep the presentation and the subsequent analysis as simple as possible, we choose a two dimensional uniform mesh. This allows us to focus on} generic\strike{futures} \blue{features} of different finite volume lattice Boltzmann methods. As shown in\strike{figure} \blue{\figref{CV}}, each cell is a square, including one standard LBM grid point,\strike{coinciding with} \blue{placed at} its center\blue{. B}oundaries of the cell are located in \blue{the} middle of\strike{to} \blue{the} standard LBM grid points.
This paper is organized as follows. Section \ref{LBM} is a review of the original lattice Boltzmann method. {In s}ections \ref{Col} and \ref{Pro},\strike{are devoted to} \blue{we} present the evolution of the velocity of each cell during a single time step, considering details of different finite volume schemes. Section \ref{MA} describes the multiscale analysis of the finite volume schemes introduced in section \ref{Pro}. Section \ref{Dis} discusses the results and concludes the paper.
\section{The Lattice Boltzmann Method} \label{LBM}
\blue{There are excellent monographs \cite{Wolf-Gladrow2000,Rothman1997,Succi2001}
and comprehensive review articles \cite{Raabe2004,Chen1998,Ladd2001}
on the lattice Boltzmann method and the related lattice-gas cellular automata (LGCA).}
Historically, the lattice Boltzmann method directly evolved from the lattice-gas cellular automata (LG\blue{C}A)~\cite{Frisch}. Particles represented by Boolean numbers in \blue{the} LG\blue{C}A are replaced by single-particle distribution functions. Consequently, the lattice Boltzmann method inherits some properties from the LGCA, including\strike{the time evolution of dynamics of particles and} the coupling between the discrete velocity space and the spatial space \blue{as well as the conservation of mass and momentum}. \blue{However, it is important to realize that, despite this historical connection between the LGCA and the LBM, the lattice Boltzmann method can also be derived in a way completely independent of the lattice-gas cellular automata, namely from the kinetic Boltzmann equation in a certain approximation \cite{HE97}.}
\strike{Here, we give a short introduction of the method. }
Similar to \blue{the} LG\blue{C}A, dynamics of evolution in \blue{the} lattice Boltzmann method\strike{consists of} \blue{can formally be divided into} two basic steps in each iteration; namely, (i) collision and (ii) propagation. \blue{As mentioned above, both mass and momentum are conserved during the collision step. This is a necessary (but not always sufficient) condition in order to recover the correct hydrodynamic behavior in the macroscopic limit. Generally, this property of the collision step is common to all mesoscopic approaches aiming at dealing with hydrodynamic phenomena such as LGCA, LBM, SRD and DPD. In the propagation step of the LGCA or the LBM, on the other hand, the} post-collision particles or distribution functions are advected entirely (without any change) along the link connecting two neighboring lattice nodes (free streaming).
\begin{figure}[h!]
\centering
\includegraphics[ width=0.25\textwidth ]{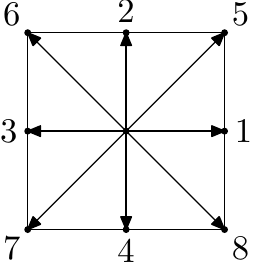}
\caption{Discrete velocities for D2Q9 model}
\label{uniformLattice}
\end{figure}
\strike{As mentioned before, this property is inherited from LGCA, but since the Boolean variables in LGCA are replaced by floating point distribution function in lattice Boltzmann, many approximations can be used to treat the uncoupled velocity and physical domain discretization \cite{dembo}.} 
In this paper, we use the \blue{two dimensional nine velocity (D2Q9) lattice Boltzmann model (see \figref{uniformLattice})}. Using the \blue{so called Bhatnagar-Gross-Krook} (BGK) approximation \blue{\cite{Bhatnagar1954}}, the equation of the evolution of the discretized distribution functions, $ F_{i}$\strike{in D2Q9 model} is\\
\begin{equation} \label{eq:evolution}
F_{i}(\bmr+ \bmc_{i}\Delta t, t+\Delta t)=F_{i}(\bmr, t)+\omega(\Fieq(\bmr, t)-F_{i}(\bmr, t))+\frac{\Delta t c_{i\alpha}}{6c^{2}}K_{\alpha}
\end{equation}
where $\Delta t$ is the time step, $\tau$ is the relaxation time and $\omega=\Delta t/\tau$. $\Fieq$ is the equilibrium distribution function, ${\bf K}=(K_{x},K_{y})$ is an applied body force which is assumed to be constant and $\alpha$ indicates different directions of physical coordinate system, $x$ and $y$. \blue{In the D2Q9 model used in the present paper,} the index $i$ refers to \blue{the} nine discrete velocities\strike{in the D2Q9 model} defined by \\
\begin{equation}
\bmc_{i} = 
\begin{cases} 
0 & i = 0 \\
(\cos[(i-1)\pi/2],\sin[(i-1)\pi/2])c & i=1, 2, 3, 4,\\
\sqrt{2}(\cos[(i-5)\pi/2+\pi/4],\sin[(i-5)\pi/2+\pi/4])c & i=5, 6, 7, 8.
\end{cases}
\end{equation}
where $c$ is defined as the ratio of the lattice spacing and the time step, $c=\Delta x/\Delta t$.\\
\blue{Physical properties of the system enter the lattice Boltzmann iteration scheme via the
quantity $\Fieq$ \equref{eq:evolution}. Obviously,
the system is 'pushed' towards $\Fieq$ with a rate $\Delta t/\tau$.
The population density $\Fieq$ is, therefore, referred to as 'equilibrium
distribution'. It is noteworthy that the term 'equilibrium'
does not refer to a global thermal equilibrium, where no flow exists.
Rather, it describes the {\it local} velocity distribution in a portion 
of fluid moving at a velocity $\bmv(\bmr)$.} Within the present lattice Boltzmann model, 
one expands $\Fieq$ in powers of the fluid velocity, $\bmv$, up to the second 
order \blue{recalling the second order expansion of the Maxwell velocity distribution.} 
This leads to
\begin{equation}\label{SLBMequi}
\Fieq(\bmr, t)=w_{i}\rho(1+\frac{3(\bmc_{i}.\bmv)}{c^{2}}+\frac{9(\bmc_{i}.\bmv)^{2}}{2c^{4}}-\frac{3v^{2}}{2c^{2}})
\end{equation}
where the weight factors $w_i$ are given by \blue{(for a derivation, see e.g.~\cite{Wolf-Gladrow2000})}
\blue{
\begin{equation}\label{weightfactors}
w_{i}=
\begin{cases}
4/9 & i=0\\
1/9 & i=1, 2, 3, 4,\\
1/36 & i=5, 6, 7, 8.
\end{cases}
\end{equation} 
}
Macroscopic quantities such \blue{as density and momentum} are\strike{estimated} \blue{obtained} locally as \blue{the zeroth and the first moments of the population density, $F_i$, respectively:}
\begin{equation}\label{rho}
\rho=\sum^{8}_{i=0}F_{i} \mbox{~~~~~~~~~~~~~~~~~and~~~~~~~~~~~~~~~~~} \rho\bmv=\sum_{i=1}^{8}\bmc_{i}F_{i}.
\end{equation}
In order to estimate momentum exchange between cells through their boundaries, the momentum flux tensor is needed. It can be defined as
\begin{equation}\label{momentumfluxtensor}
\Pi_{\alpha\beta}=\sum_{i=1}^{8}\bmc_{i \alpha} \bmc_{i \beta} F_i .
\end{equation}
In the above equation, $\alpha$ and $\beta$ indicate different directions of Cartesian coordinate system, $x$ and $y$.
\blue{Note that, since $\bmc_0=\bm{0}$, the term with $i=0$ has no effect on the fluid velocity as well as the momentum flux tensor.}
It is of common practice to decompose equation (\ref{eq:evolution}) into two\strike{equations} \blue{parts}, the first\strike{one resembles} \blue{part is the so called} collision \blue{step} and the second one implements \blue{the free} propagation \blue{(streaming)} of the distribution functions between two neighboring lattice nodes directly linked via the velocity vector $\bmc_i$. \varnik{The collision step is considered to occur within a time interval which is infinitesimally small compared to all relevant times in the problem. We therefore consider it as an instantaneous process. The streaming step, on the other hand, happens in the finite time interval $[t, t+\Delta t]$}.
\begin{eqnarray}\label{decomposed}
F^*_{i}(\bmr, t)&=&F_{i}(\bmr, t)+\omega(F_{i}^{eq}(\bmr, t)-F_{i}(\bmr, t))+\frac{\Delta t c_{i\alpha}}{6c^{2}}K_{\alpha}\\
F_{i}(\bmr+ \bmc_{i}\Delta t, t+\Delta t)&=&F^*_{i}(\bmr, t).
\end{eqnarray}
It can be easily shown that\strike{the collision step is mass and momentum conserving:}
\blue{the requirement of mass and momentum conservation during the collision step is equivalent to the following two conditions on the moments of the distribution function, $F_i$, namely}
\begin{equation}\label{Consevartion}
\sum_{i=0}^{8} F_{i}(\bmr, t) = \sum_{i=0}^{8}\Fieq(\bmr, t) \mbox{~~~~~~~~~~~~ and ~~~~~~~~~~~~}
\sum_{i=1}^{8} F_{i}(\bmr, t)\bmc_i = \sum_{i=1}^{8}\Fieq(\bmr, t)\bmc_i.
\end{equation}
In the next two sections, we show how the velocity of each cell changes during these two steps in \blue{a single iteration cycle}.
\section{Finite Volume Analysis of the Momentum Change Due to Collision} \label{Col}
\blue{Although the} collision step is modified differently in each finite volume lattice Boltzmann scheme, \strike{but}\blue{the} mass and momentum conserving properties of \blue{the} standard lattice Boltzmann \blue{method} are not violated in any of those schemes. Consequently, the collision step\strike{will} blue{does} not change the momentum of the cells. \HodaRemove{If external forces are applied, the velocity of the center of mass of each cell
}\Hoda{If an external field is applied, for each cell the velocity of the center of mass} will increase by a constant when the external field is constant.
\section{Finite Volume Analysis of the Momentum Change Due to Propagation} \label{Pro}
After the collision step, the system undergoes a propagation step,\strike{ and reaches its final state of the iteration.}
during which the momentum of cells can change due to the exchange of particles.
\strike{, which is equivalent to movement of distribution function in lattice Boltzmann method.}
\blue{However, in the standard lattice Boltzmann method, populations, $F_i$, are advected entirely 
(i.e.~with no loss or gain) between two neighboring lattice nodes linked along the velocity vector $\bmc_i$. 
In contrast to this, within a finite volume approach, one first determines the total mass flux as well as the momentum 
flux through the cell's boundaries by integrating the component of the corresponding current projected onto the direction
{\it normal} to the cell's surface. More formally, and considering the case of momentum exchange, one computes 
the total change in momentum of a cell via} 
\begin{equation} \label{P}
\Delta \bm{P}=\Delta t \sum_{i=1}^{8} \int F_i \bmc_{i} [\bmc_{i} \blue{\cdot} d\bmS ].
\end{equation}
Here, the integral is over the whole boundary of a cell. For the simple case of a regular \blue{rectangular} lattice, \blue{each node can be surrounded by a rectangular finite volume cell. In this case,} the above integral reduces to
\begin{eqnarray} 
(\frac{\Delta P}{\Delta t})_{\alpha} &=& 
\green{\sum_{i=1}^{8}} \int F_i c_{i \alpha} [\bmc_{i} \cdot d\bmS ] \\ \label{DeltaP1}
&=& \green{\sum_{i=1}^{8}} c_{i \alpha} c_{i x} [ \Fiav(\bmr + \frac{\Delta x}{2} \bmi ) - \Fiav(\bmr - \frac{\Delta x}{2}\bmi ) ] \Delta y \\ \nonumber
&+& \green{\sum_{i=1}^{8}} c_{i \alpha} c_{i y} [\Fiav(\bmr+\frac{\Delta y}{2} \bmj )-\Fiav(\bmr-\frac{\Delta y}{2} \bmj )] \Delta x .
\end{eqnarray}
where $\Fiav$\strike{at any position} is the average of $F_i$ over the corresponding edge of the cell. \referee{If the edge is horizontal, $\Fiav(\bmr)$ is defined as 
\begin{equation}\label{def_f_ave_x}
\Fiav(\bmr)=\frac{1}{\Delta x}\int_{\bmr - \frac{\Delta x}{2} \bmi}^{\bmr + \frac{\Delta x}{2} \bmi} F_i(\mathcal{\\R}) d\mathcal{\\R}
\end{equation}
and it is \varnik{given by}
\begin{equation}\label{def_f_ave_y}
\Fiav(\bmr)=\frac{1}{\Delta y}\int_{\bmr - \frac{\Delta y}{2} \bmj}^{\bmr + \frac{\Delta y}{2} \bmj} F_i(\mathcal{\\R}) d\mathcal{\\R}
\end{equation}
\varnik{if} the edge is vertical.
}\green{The first two terms in \equref{DeltaP1} arise from fluxes across the right and left boundaries ($d\bmS=(\Delta y, 0)$) whereas the terms proportional to $\Delta x$ represent the contribution of the flux through the top and bottom edges of the cell ($d\bmS=(0, \Delta x)$).}
\green{In order to simplify the notation, we make use of Einstein summation convention over repeated indices and rewrite \equref{DeltaP1} as}
\begin{eqnarray} \label{DeltaP}
(\frac{\Delta P}{\Delta t})_{\alpha}&=& c_{i \alpha} [ \Fiav(\bmr + \frac{c_{i x} \Delta t}{2}\bmi ) - \Fiav(\bmr-\frac{c_{i x}\Delta t}{2}\bmi )] c \Delta y \\ \nonumber 
&+& c_{i \alpha} [\Fiav(\bmr+\frac{c_{i y}\Delta t}{2}\bmj)-\Fiav(\bmr-\frac{c_{i y}\Delta t}{2}\bmj)] c \Delta x
\end{eqnarray}
\hide{
\red{Nima: Why do you set $c$ for $c_{i x}$ and $c_{i y}$? I would keep $c_{i x}\Delta y$ as well as $c_{i y}\Delta x$ as in \equref{DeltaP1}}}
\blue{Thus, for the simple case of a square lattice studied in this work, the computation of the total momentum change within a finite volume cell requires the knowledge of the population density, $F_i$, midway between the node surrounded by the cell and 
the lattice nodes linked to it via a velocity vector $\bmc_i$. Since only the values of $F_i$ {\it at} the lattice nodes are exactly known, some approximation is necessary in order to obtain the required quantities. Therefore, we discuss}
in the next subsection the details of different methods of evaluating $\Fiav$, including \blue{the} central scheme \blue{and} constant \blue{as well as} linear upwind schemes\strike{will be discussed} \cite{Steibler}. \blue{By conducting} a full Chapman-Enskog analysis of these finite volume schemes in section \ref{MA}, \blue{we also study the resulting macroscopic equations and, in particular, the relation between the relaxation time, $\tau$, and the kinematic viscosity, $\nu$, within each approximate method}.
\subsection{Computation of the Fluxes}
\blue{As mentioned above,} in order to estimate fluxes, $\Fiav$ should be evaluated at the cell faces. In the context of finite volume methods, there are many different approaches for flux evaluation \cite{Peric}. \referee{In 1959 Godunov proposed a scheme in which flux evaluation is based on the value of conserved variables which are considered to be piecewise constant over the cells \cite{Godunov59}. Later in 1979, van Leer extended the Godunov's scheme to a higher order total variation diminishing (TVD) method, through use of  piecewise linear approximation within each cell and flux limiters \cite{VanLeer79}. Methods introduced in the finite volume method to solve partial differential equations are implemented in the finite volume lattice Boltzmann method to solve the Boltzmann equation \varnik{(see e.g.\ \cite{Succi,Nanneli,UbertiniSucci,Peng,Xi_PRE})}.} In this paper, two different upwind schemes introduced by Steibler et al. \cite{Steibler} and one central scheme introduced by Peng et al. \cite{Peng} are analyzed.
\subsubsection{Constant Upwind Scheme }
A very simple upwind scheme results from the following approximation 
\begin{equation} \label{upwindconstdefinitionX1}
\Fiav(\bmr+\frac{\Delta x}{2}\bmi) \simeq F_i(\bmr) \mbox{~~~~~~~~~~~~ and ~~~~~~~~~~~~} 
\Fiav(\bmr-\frac{\Delta x}{2}\bmi) \simeq F_i(\bmr-\Delta x\bmi)
\end{equation}
\begin{equation} \label{upwindconstdefinitionY1}
\Fiav(\bmr+\frac{\Delta y}{2}\bmj) \simeq F_i(\bmr)
\mbox{~~~~~~~~~~~~ and ~~~~~~~~~~~~}
\Fiav(\bmr-\frac{\Delta y}{2}\bmj) \simeq F_i(\bmr-\Delta y\bmj)
\end{equation}
\\
\referee{This scheme is the basic Godunov scheme, which simply assumes constant approximations for the value of variables and results in the first order upwind discretization.}
\green{Inserting} equations (\ref{upwindconstdefinitionX1}) and (\ref{upwindconstdefinitionY1}) into 
\equref{DeltaP}, $({\Delta P}/{\Delta t})_{\alpha}$ can be estimated as
\begin{flalign}
(\frac{\Delta P}{\Delta t})_{\alpha}&=c_{i \alpha} [F_i(\bmr) -F_i(\bmr-c_{i x}\Delta t\bmi) +F_i(\bmr) - F_i(\bmr-c_{i y}\Delta t\bmj) ] c \Delta x=\\ 
& =c_{i \alpha} [ c_{i\beta}\Delta t \partial_{\beta} F_i \vert_{\bmr}-\frac{c_{i\beta}c_{i\gamma}(\Delta t)^2}{2}\partial_{\beta}\partial_{\gamma} F_i \vert_{\bmr} +O((\Delta x)^3)] c \Delta x \label{UpwindConsA}
\end{flalign}
It is worth mentioning that, in contrast to central and linear upwind schemes (see equations (\ref{centeralA}) and (\ref{upwindLineA}) ), terms of the second order do not vanish here. As will be shown in section \ref{MA} and appendix A, this gives rise to a finite numerical viscosity within the constant upwind approach, an artifact absent both in the central and \HodaRemove{the }linear upwind schemes.
\subsubsection{Central Scheme}
Average value of $F_i$ over each face of the cell can be determined by averaging between distribution functions at the both ends of the corresponding face but end points of the cells are not coinciding with the lattice nodes so the values of $F_{i}$ are not known and should be approximated. Each end point has four neighboring lattice nodes and values of $F_{i}$s at these point can be used to interpolate the value of distribution functions at the corresponding end point.

\referee{This method is equivalent to the extension of Godunov's method to use piecewise linear approximation of each cell. Although this method results in second order accuracy in space, it introduces unphysical oscillations in problems with shocks or discontinuities. These spurious oscillations are well-known for higher order schemes (e.g.\ the Lax-Wendroff scheme \cite{Lax} and Beam-Warming scheme \cite{Beam}). These numerical oscillations are due to the non-monotonicity of the higher order schemes\cite{Hirsch}. In order to deal with this problem, the concept of flux limiters is introduced into the context of CFD. A systematic analysis and derivation of a class of flux limiters for high resolution TVD second order schemes is done in \cite{Sweby}. Normally, the need for  flux limiters arise when there are discontinuities in the problem and in this paper for simplicity we focus on problems with smooth solutions.} \referee{For such a smooth problem,} it can be easily shown that \\
\begin{equation} \label{centralschemedifinitionX}
\Fiav(r\pm\frac{c_{i x}\Delta t}{2}\bmi) \simeq \frac{F_i(r\pm c_{i x}\Delta t\bmi)+ F_i(r)}{2}
\end{equation}
\begin{equation} \label{centralschemedifinitionY}
\Fiav(r\pm\frac{c_{i y}\Delta t}{2}\bmj) \simeq \frac{F_i(r\pm c_{i y}\Delta t\bmj)+ F_i(r)}{2}
\end{equation}
Substituting equation (\ref{centralschemedifinitionX}), as an approximation of $\Fiav$, in the first term of equation (\ref{DeltaP}) leads to 
\begin{flalign} 
& \Fiav(r+\frac{c_{i x}\Delta t}{2}\bmi)-\Fiav(\bmr-\frac{c_{i x}\Delta t}{2}\bmi)=\\
&=\frac{F_i(\bmr+c_{i x}\Delta t\bmi)+F_i(\bmr)}{2}-\frac{F_i(\bmr-c_{i x}\Delta t\bmi)+F_i(\bmr)}{2}=\\
&=\frac{F_i(\bmr+c_{i x}\Delta t\bmi)-F_i(\bmr-c_{i x}\Delta t\bmi)}{2}=\\
&={c_{i x}\Delta t}\partial_{x} F_i \vert_{\bmr}+O((\Delta x)^3) \label{centeralschemeAx}
\end{flalign}
It must be emphasized that terms related to second derivatives of $F_i$s will cancel each other and despite the fact that only first derivatives of $F_i$s appear in equation (\ref{centeralschemeAx}), the first non-vanishing error term is of the order $O((\Delta x)^3)$. \referee{Generally, this is not the case when the geometry of the cell is more complicated.}
\green{Similarly}, the second term in \equref{DeltaP} can be simplified according to \equref{centralschemedifinitionY},
\begin{equation} 
\Fiav(\bmr+\frac{\Delta y}{2}\bmj)-\Fiav(\bmr-\frac{\Delta y}{2}\bmj)
={c_{i y}\Delta t}\partial_{y} F_i \vert_{\bmr}+O((\Delta x)^3) \label{centeralschemeAy}
\end{equation}
\\
Substituting equations (\ref{centeralschemeAx}) and (\ref{centeralschemeAy}) in \equref{DeltaP} 
results in 
\begin{equation} \label{centeralA}
(\frac{\Delta P}{\Delta t})_{\alpha}=c_{i \alpha} [{c_{i \beta}\Delta t}\partial_{\beta} F_i \vert_{\bmr} + O(\Delta x)^3] c\Delta x
\end{equation}
\hide{
\red{Nima: I think that here $c_{i \beta}$ must appear with a power of two; see \equref{DeltaP}.
Please check for similar aspects in all the equations below.}
}
\subsubsection{Linear Upwind Scheme}
A more accurate upwind scheme can be constructed using the Taylor expansion, in the sense that $F_{i}$ at cell's boundaries can be approximated with
\begin{equation} \label{upwindlineardefinitionplusX}
\Fiav(\bmr+\frac{c_{i x}\Delta t}{2}\bmi) \simeq F_i(\bmr)+\frac{c_{i x}\Delta t}{2}\partial^1_{x} F_i \vert_{\bmr}+O((\Delta x)^2)
\end{equation}
\begin{equation} \label{upwindlineardefinitionminusX}
\Fiav(\bmr-\frac{c_{i x}\Delta t}{2}\bmi) \simeq F_i(\bmr-c_{i x}\Delta t \bmi)+\frac{c_{i x}\Delta t}{2}\partial^1_{x} F_i \vert_{\bmr}+O((\Delta x)^2)
\end{equation}
\begin{equation} \label{upwindlineardefinitionplusY}
\Fiav(\bmr+\frac{c_{i y}\Delta t}{2}\bmj) \simeq F_i(\bmr)+\frac{c_{i y}\Delta t}{2}\partial^1_{y} F_i \vert_{\bmr}+O((\Delta x)^2)
\end{equation}
\begin{equation} \label{upwindlineardefinitionminusY}
\Fiav(\bmr-\frac{c_{i y}\Delta t}{2}\bmj) \simeq F_i(\bmr-c_{i y}\Delta t \bmj)+\frac{c_{i y}\Delta t}{2}\partial^1_{y} F_yi \vert_{\bmr}+O((\Delta x)^2)
\end{equation}
Substituting equations (\ref{upwindlineardefinitionplusX}) and (\ref{upwindlineardefinitionminusX}) in the first term of \equref{DeltaP} results in,
\begin{flalign} \label{upwindAx}
& \Fiav(\bmr+\frac{c_{i x}\Delta t}{2}\bmi) - \Fiav(\bmr-\frac{c_{i x}\Delta t}{2}\bmi)=\\
&=F_i(\bmr)+\frac{c_{i x}\Delta t}{2}\partial_{x} F_i \vert_{\bmr}-(F_i(\bmr-c_{i x}\Delta t)+\frac{c_{i x}\Delta t}{2}\partial^1_{x} F_i \vert_{r-c_{i x}\Delta t\bmi})=\\
&=F_i(\bmr)-F_i(\bmr-c_{i x}\Delta t)+\frac{c_{i x}\Delta t}{2}\partial_{x} F_i \vert_{\bmr}-\frac{c_{i x}\Delta t}{2}\partial^1_{x} F_i\vert_{r-c_{i x}\Delta t\bmi}=\\
&=c_{i\beta}\Delta t \partial_{\beta} F_i \vert_{\bmr}-\frac{c_{i\beta}c_{i\gamma}(\Delta t)^2}{2}\partial_{\beta}\partial_{\gamma} F_i \vert_{\bmr} +\frac{c_{i x}c_{i\gamma}(\Delta t)^2}{2}\partial_{x}\partial_{\gamma} F_i \vert_{\bmr}+O((\Delta x)^3)=\\
&=c_{i x}\Delta t \partial_{x} F_i \vert_{\bmr}+O((\Delta x)^3)
\end{flalign}
Using equations (\ref{upwindlineardefinitionplusY}) and (\ref{upwindlineardefinitionminusY}), the same procedure can be applied for the second term of equation (\ref{DeltaP}). Combining both terms of equation (\ref{DeltaP}),
\begin{equation} \label{upwindLineA}
(\frac{\Delta P}{\Delta t})_{\alpha}=c_{i \alpha} [ {c_{i \beta}\Delta t}\partial_{\beta} F_i \vert_{\bmr} +O((\Delta x)^3) ] c \Delta x
\end{equation}
which is identical to equation (\ref{centeralA}). 
\section{Multiscale Analysis} \label{MA}
In this section we introduce a multi scale expansion of the distribution function around the equilibrium distribution function \cite{Wolf-Gladrow2000}
\begin{equation} \label{Enskog_Expansion}
F_i=F^0_i+\epsilon F^1_i+ \epsilon^2 F^2_i+O(\epsilon^3)
\end{equation}
and the derivatives
\begin{equation} \label{expansion of derivatives}
\partial_{x}=\epsilon \partial^1_x 
\end{equation}
into different schemes discussed above. The first term is the equilibrium distribution function ($F_i^0$) and the first non-equilibrium term of equation (\ref{Enskog_Expansion}), $F^1_i$, can be approximated \referee{(for details refer to appendix A)} as
\begin{equation} \label{F1}
F^1_i=-\frac{\Delta t}{\omega} c_{i\gamma} \partial^1_{\gamma} F^0_i + \frac{\Delta t}{6c^2\omega} c_{i\gamma} K_{\gamma}
\end{equation}
where ${\bf K}=(K_{x},K_{y})$ is the external force.
\subsection{Central and Linear Upwind Schemes}
Equations (\ref{centeralA}) and (\ref{upwindLineA}) are identical. The Chapman-Enskog analysis of equation (\ref{centeralA}) is done here and the result is also valid for equation (\ref{upwindLineA}).\\
Substituting equations (\ref{Enskog_Expansion}), (\ref{expansion of derivatives}) and (\ref{F1}) in (\ref{centeralA}) and keeping the terms up to $\epsilon^2$ results in
\begin{equation} \label{EnskogCen}
\frac{\Delta \bf{P}}{ \Delta t} =\epsilon \bmc_{i} c_{i\beta} \partial^1_{\beta} F^0_i (\Delta x)^{2}+\epsilon^2 \bmc_{i} c_{i\beta} \partial^1_{\beta} F^1_i (\Delta x)^{2}+O(\epsilon^3) 
\end{equation}
\begin{equation} \label{Ai_FinalC}
\frac{\Delta \bf{P}}{ \Delta t} \simeq \epsilon \bmc_{i } c_{i\beta} \partial^1_{\beta} F^0_i (\Delta x)^{2}+ \epsilon^2 \bmc_{i }(\frac{-1}{\omega})c_{i\beta}c_{i\gamma}\Delta t\partial^1_{\beta}\partial^1_{\gamma} F^0_i (\Delta x)^{2}
\end{equation}
\subsection{Constant Upwind Scheme }
Substituting equations (\ref{Enskog_Expansion}), (\ref{expansion of derivatives}) and (\ref{F1}) in (\ref{UpwindConsA}) and keeping the terms up to $\epsilon^2$ results in
\begin{equation} \label{EnskogUpC}
\frac{\Delta \bf{P}}{ \Delta t}=\epsilon \bmc_{i} c_{i\beta} \partial^1_{\beta} F^0_i (\Delta x)^{2}+\epsilon^2 \bmc_{i } c_{i\beta} \partial^1_{\beta} F^1_i (\Delta x)^{2}-\epsilon^2 \bmc_{i } \frac{c_{i\beta}c_{i\gamma}\Delta t}{2}\partial^1_{\beta}\partial^1_{\gamma} F^0_i (\Delta x)^{2} + O(\epsilon^3) 
\end{equation}
\begin{equation} \label{Ai_FinalUpC}
\frac{\Delta \bf{P}}{ \Delta t} \simeq \epsilon \bmc_{i } c_{i\beta}\partial^1_{\beta} F^0_i (\Delta x)^{2}- \epsilon^2 (\frac{1}{\omega}+\frac{1}{2})\bmc_{i } c_{i\beta}c_{i\gamma}\Delta t\partial^1_{\beta}\partial^1_{\gamma} F^0_i (\Delta x)^{2}
\end{equation}
\subsection{Recovering Navier-Stokes Equation}
The first velocity moment of $\epsilon$ and $\epsilon^2$ terms of (\ref{Ai_FinalC}) and (\ref{Ai_FinalUpC}), reproduce the Navier-Stokes equation. In Appendix A details of the calculations for central scheme and constant upwind scheme are explained. It can be seen that the  kinematic viscosity, $\nu$, will be different among the different schemes:
\begin{equation}
\nu = 
\begin{cases} 
c_{s}^2 \tau & \text{central, linear upwind}\\
c_{s}^2 (\tau+\Delta t/2) & \text{constant upwind} 
\end{cases}
\end{equation}
where $c_{s}= c/\sqrt{3}$ is the sound speed.
\section{Discussion} \label{Dis}
As shown here, different approaches for solving the Boltzmann equation may result in different values for the viscosity. The viscosity estimated from the Chapman-Enskog analysis of the Boltzmann equation with the BGK approximation is exactly $c^2_s \tau$ and the difference between $c^2_s \tau$ and the viscosity obtained from the numerical scheme is referred to as numerical viscosity. The numerical viscosity is an artifact of the solution method and generally undesirable. Through the Chapman-Enskog analysis we showed that the numerical viscosity \green{identically vanishes both within the} central scheme and \green{the} linear upwind schemes. \green{The constant upwind scheme, on the other hand, yields a finite numerical viscosity of $c^2_s \Delta t/2$. Even though undesirable, this numerical viscosity} admits the positivity of the numerical viscosity as reported in \cite{Steibler}. \\
Generally speaking, regarding the accuracy, some finite difference schemes and finite volume schemes will result in the same set of discrete equations in \green{the} limit of a uniform lattice. Comparison of constant upwind scheme and central scheme in present work with their finite difference lattice Boltzmann equivalents shows agreement in value of viscosity \cite{Sofonea}. It is typical of first order schemes (e.g. constant upwind scheme; see equation (\ref{UpwindConsA}) and the comment on it) to show higher viscosity in comparison to second order schemes (e.g. central scheme), which are expected to show oscillations in the solution \cite{Peric} . \\
\newstrike{The argument about the rule of collision can be used to extend the validity of our analysis to any reasonable collision rule while fulfilling mass and momentum conservation properties[{\it Is it really ANY collision rule? the word ``reasonable'' is not clear at all for me! }].\\}
\green{In summary,} we introduced a systematic procedure, which enables further analysis of the finite volume lattice Boltzmann methods \green{and presented a derivation of the relation between the kinematic viscosity of the fluid and the lattice Boltzmann relaxation parameter for some typical approximate schemes for the flux evaluation within the finite volume approach to the lattice Boltzmann method.} The next step in this framework \green{could} be the analytical study of different boundary conditions, the effect of boundary condition on the stress tensor, slip as an artifact in mesoscopic methods \referee{and analysis of the effect of the more complex cell geometries on the properties of different schemes.}
\appendix
\section{Chapman-Enskog Expansion}
The Chapman-Enskog expansion is a method to derive the Navier-Stokes equation and its transport coefficients from the Boltzmann equation. This method has been developed by Chapman and Enskog \cite{ChapmanOriginal,Enskog}.
\referee{Here we just consider the Chapman-Enskog analysis for \varnik{two dimensional nine velocity (D2Q9)} lattice Boltzmann model}. The distribution functions $F_i(x,t)$ are expanded around the equilibrium distributions $F^0_i(x,t)$
\begin{equation} \label{App_Enskog_Expansion}
F_i(x,t)=F^0_i(x,t)+\epsilon F^1_i(x,t)+ \epsilon^2 F^2_i(x,t)+O(\epsilon^3)
\end{equation}
with 
\begin{eqnarray}
\sum_{i=0}^{8}\F{1}=0 \mbox{~~~~~~~~~~~~ and ~~~~~~~~~~~~} \sum_{i=1}^{8}\bmc_{i}\F{1}=0,\\
\sum_{i=0}^{8}\F{2}=0 \mbox{~~~~~~~~~~~~ and ~~~~~~~~~~~~} \sum_{i=1}^{8}\bmc_{i}\F{2}=0.
\end{eqnarray}
The small expansion parameter $\epsilon$ \HodaRemove{can be considered to be}\Hoda{might be considered as} \begin{enumerate}
\item the Knudsen number which is the ratio between the mean free path and the characteristic length scale of the flow.
\item a formal parameter in the expansions which allows one to keep track of the
relative orders of magnitude of the various terms. It will be considered
only as a label and will be dropped out of the final results by setting $\epsilon=1$.
\end{enumerate}
As an example consider the expansion $F_i(x,t)=F^0_i(x,t)+\epsilon F^1_i(x,t)$. In discussions one may
consider $F^0_i(x,t)$ and $F^1_i(x,t)$ as quantities of the same order of magnitude and argue
that the second term of the expansion is small because $\epsilon$ is a small quantity
whereas in the formal calculations $F^1_i(x,t)$ is small compared to $F^0_i(x,t)$ and $\epsilon$ is
only a label to keep track of the relative size of the various terms. The $\epsilon$ in
this second sense can be set equal to one after finishing all transformations \Hoda{\cite{Wolf-Gladrow2000}}.
As the last point about equation (\ref{App_Enskog_Expansion}), one should mention that the series resulting from the
Chapman-Enskog procedure is probably not convergent but asymptotic~\referee{\cite{Santos}}.\\
Considering the steady state case, only spatial changes of $F_i$ are taken into account and  temporal derivations of the distribution functions will vanish.
\begin{eqnarray}\label{app_steady}
F_{i}(\bmr+ \bmc_{i}\Delta t, t+\Delta t)&=&F_{i}(\bmr+ \bmc_{i}\Delta t, t)
\end{eqnarray}
So equation (\ref{eq:evolution}) will be 
\begin{equation} \label{eq:app_evolution}
F_{i}(\bmr+ \bmc_{i}\Delta t,t)=F_{i}(\bmr, t)+\omega(\Fieq(\bmr, t)-F_{i}(\bmr, t))+\frac{\Delta t c_{i\alpha}}{6c^{2}}K_{\alpha}
\end{equation}
One can expand\HodaRemove{ed} the left hand of the equation (\ref{eq:app_evolution}) into a Taylor series up to terms of second order as follows 
\begin{eqnarray}\label{app_taylor expansion}
F_{i}(\bmr+ \bmc_{i}\Delta t,t)=F_{i}(\bmr, t)+c_{i\alpha}\Delta t \rond{}{\alpha}{}+\frac{(\Delta t)^2}{2} c_{i\alpha}c_{i\beta}\rondDo{}{\alpha}{\beta}{}
\end{eqnarray}
Here one spatial scale with the following scaling will be introduced
\begin{equation} \label{app_scaling}
\partial_{\alpha}=\epsilon \partial^{1}_{\alpha}
\end{equation}
Substitut\green{ion} of expansions (\ref{App_Enskog_Expansion}) and (\ref{app_scaling}) and \equref{app_taylor expansion} into equation (\ref{eq:app_evolution}) leads to
\begin{eqnarray} \label{app_firstOrder}
0&=& \epsilon( c_{i\gamma} \rond{1}{\gamma}{0}+\frac{\omega}{\Delta t} \F{1}-\frac{c_{i\gamma} }{6c^2} K_{\gamma} )+\\ \label{app_secondOrder}
&& \epsilon^{2} (c_{i\gamma} \rond{1}{\gamma}{1}+\frac{\Delta t}{2}\rondDo{1}{\beta}{\gamma}{0}+\frac{\omega}{\Delta t} \F{2}).
\end{eqnarray}
\green{Setting} terms of first order in $\epsilon$ \green{to zero} result in
\begin{equation}\label{F1_App}
\F{1}=-\frac{\Delta t}{\omega} c_{i\gamma} \rond{1}{\gamma}{0} + \frac{\Delta t}{6c^2\omega} c_{i\gamma} K_{\gamma}
\end{equation}
Using the above equation, it is possible to calculate terms of order $\epsilon$ and $\epsilon^{2}$ \referee{for the central and linear upwind schemes} in \varnik{equation} (\ref{Ai_FinalC}) \referee{through substituting equation (\ref{F1_App}) in equation (\ref{EnskogCen})}. The same procedure can be applied to terms of \varnik{equation} (\ref{EnskogUpC}) \referee{in constant upwind scheme which results in \varnik{equation} (\ref{Ai_FinalUpC})}.
\\
\referee{Terms of order $\epsilon$ in the central and linear upwind and also constant upwind schemes are identical (refer to equations (\ref{Ai_FinalC}) and (\ref{Ai_FinalUpC})). These terms }lead to \varnik{the} steady state Navier-Stokes equation without \green{viscous} friction (\green{Euler equation}) \newstrike{in incompressible limit}
\begin{eqnarray}\label{appFirst}
\sum_{i=1}^{8} \bmc_{i} ( c_{i\beta}\Delta t\partial^1_{\beta} \F{0}-\frac{c_{i\beta} }{6c^2} K_{\beta})&=&\rho \bmv.\nabla\bmv+{\nabla p}-{\bf K} 
\end{eqnarray}
where $p=\frac{\rho}{c^{2}_{s}}$.\\
\referee{Considering the central and linear upwind schemes,} the second term in equation (\ref{Ai_FinalC}) in incompressible limit results in viscous term of the Navier-Stokes equation. 
\begin{equation}\label{appSecond}
\sum_{i=1}^{8} \bmc_{i }(\frac{-1}{\omega})c_{i\beta}c_{i\gamma}(\Delta t)^2\partial^1_{\beta}\partial^1_{\gamma} F^0_i={\Delta t}{c^{2}_{s}} (\frac{-1}{\omega}) \rho\nabla^{2}\bmv
\end{equation}
Summation of \varnik{equations (\ref{appFirst}) and (\ref{appSecond}) results in the} steady state Navier-Stokes equation in incompressible limit \referee{for the central and linear upwind approaches},
\begin{equation}\label{NS}
\mu \nabla^{2}\bmv-\rho \bmv.\nabla\bmv-{\nabla p}+{ \bf K}=0
\end{equation}
where $\mu=\rho c^2_s \tau $ is the dynamic viscosity.
\referee{Comparing the second order terms in equations (\ref{Ai_FinalC}) and (\ref{Ai_FinalUpC}), it is clear that the last term in constant upwind scheme doesn't appear in the central or linear upwind methods. Contribution of this term is
\begin{equation}\label{extra}
\sum_{i=1}^{8} \bmc_{i }(\frac{-c_{i\beta}^2\Delta t}{2})\partial^1_{\beta}\partial^1_{\beta} F^0_i={\Delta t}{c^{2}_{s}} (\frac{-1}{2}\rho\nabla^{2}\bmv-\rho\partial_x\partial_y\bmv).
\end{equation}
This terms adds up with the viscous term resulting from equation (\ref{appSecond}) and leads to the viscosity of the constant upwind scheme, $\mu=\rho c^2_s (\tau+\frac{1}{2})$, which is higher than the viscosity in central and upwind schemes.}
\section*{Acknowledgments}
Financial support from the Deutsche Forschungsgemeinschaft (German Research Association) through grant GSC 111 is gratefully acknowledged.

\bibliographystyle{aipnum4-1.bst}


\end{document}